\documentclass[]{aastex631}
\shorttitle{Evidence for the percentric passage of the Sagittarius dwarf galaxy}
\shortauthors{Wang C. et al.}
\graphicspath{{./}{figures/}}

\begin{document}

\title{A North-South Metallicity Asymmetry in the Outer Galactic disk -- Evidence for the Pericentric Passage of the Sagittarius Dwarf Galaxy}

\author[0000-0002-8713-2334]{Chun Wang}
\altaffiliation{Corresponding authors (wchun@tjnu.edu.cn; huangyang@ucas.ac.cn)}
\affiliation{Tianjin Astrophysics Center, Tianjin Normal University, Tianjin 300387, People's Republic of China.}

\author[0000-0003-3250-2876]{Yang Huang}
\altaffiliation{Corresponding authors (wchun@tjnu.edu.cn; huangyang@ucas.ac.cn)}
\affiliation{School of Astronomy and Space Science, University of Chinese Academy of Sciences, Beijing 100049,  People's Republic of China.} 
\affiliation{National Astronomical Observatories, Chinese Academy of Sciences, Beijing 100012, P.R. People’s Republic of China.}

\author[0000-0003-4573-6233]{Timothy C. Beers}
\affiliation{Department of Physics and Astronomy and JINA Center for the Evolution of the Elements (JINA-CEE), University of Notre Dame, Notre Dame, IN 46556, USA}

\author{Payel Das}
\affiliation{Astrophysics Research Group, University of Surrey, Guildford, Surrey, GU2 7XH.}

\author[0000-0003-2471-2363]{Haibo Yuan}
\affiliation{School of Physics and Astronomy, Beijing Normal University, Beijing 100875, People’s Republic of China.}

\author[0000-0003-3864-068X]{Lizhi Xie}
\affiliation{Tianjin Astrophysics Center, Tianjin Normal University, Tianjin 300387, People's Republic of China.}

\author{Shi Shao}
\affiliation{National Astronomical Observatories, Chinese Academy of Sciences, Beijing 100012, P.R. People’s Republic of China.}



\begin{abstract}
We present maps of the mean metallicity distributions on the Galactocentric $R$--$Z$ plane at different azimuthal angles using red clump 
stars selected from the LAMOST and APOGEE surveys. In the inner disk ($R < $ 11\,kpc), the metallicity distribution is symmetric between the upper and lower disk. However, we find a North-South metallicity asymmetry in the outer disk ($R > 11$\,kpc), especially towards the anti-Galactic center ($-5^\circ < \Phi < 15^\circ$) direction.  By further dissecting the map in age space, we detect this asymmetry across all mono-age stellar populations.  However, the asymmetry is less pronounced in older populations ($\tau > 8$ Gyr) compared to younger ones ($\tau < 6$\,Gyr). This reduced significance likely stems from three factors: larger age uncertainties, fewer stars in the outer disk, and the kinematically hotter nature of older populations. 
The observed metallicity asymmetry may be the consequence of the purturbation of  the recent pericentric passage through the Galactic disk and tidal force of the well-known Sagittarius dwarf galaxy. 
\end{abstract}



\section{Introduction}

It is well known that the Milky Way disk is neither temporally constant  nor axisymmetric.   In recent decades, several significant non-steady and non-axisymmetric observational features of the Galactic disk have been discovered. These include the North-South asymmetry in the stellar number density and velocity 
\citep{Gomez2012,Widrow2012, carlin2013,williams2013}, moving groups \citep{Dehnen1999, Antoja2008}, the Galactic disk warp \citep{Binney1992, Levine2006, Chen2019, Poggio2020, Huang2024}, and the spiral feature in the $Z$-$V_{Z}$ phase space \citep{Antoja2018, Tian2018, Laporte2019, Wang2019, Bland-Hawthorn2021}. 
These kinematic and structural features result from a complex interplay of various physical processes,  including the inflow and outflow of gas \citep{Larson,Colavitti2008,Pezzulli2016,Andrews2017,Toyouchi2018,Grisoni2018}, the feedback from supernovae \citep{Kobayashi}, secular evolution induced by non-axisymmetric perturbations arising from structures such as the central bar or spiral arms \citep{sellwood,Roskar2008,Schonrich,Loebman2011,Haywood2016}, and the accretion
of dwarf galaxies \citep{Quinn1993, Velazquez1999, Villalobos2008, Gomez2013}. Among these processes, the pericenter passages through the Galactic disk of the Sagittarius dwarf galaxy (Sgr) are thought to play an important role \citep{Purcell2011, Laporte2018}.

According to observational studies and orbit simulations, Sgr has had three pericentric passages through the Galactic disk: at approximately 5.7\,Gyr, 1.9\,Gyr, and 1\,Gyr ago \citep{Ruiz2020}.  The three pericentric passages play pivotal roles in reconstructing the Galactic disk and produce some structural, chemical and kinematic features.  Although the first passage of the Sagittarius dwarf galaxy (Sgr) occurred 5.7\,Gyr ago, it may also have resulted in some imprints, as it was much more massive ($\sim 6 \times 10M_{\odot}$) during its first passage compared to the latter ones \citep{Laporte2018}.   These include overdensities such as the Monoceros Ring \citep{Newberg2002} and the Anti-Center Stream \citep{Grillmair2006,Laporte2019,Laporte2020},  the vertical-velocity dispersion jump at 6\,Gyr \citep{Das2024} .  The most recent passage of Sgr is believed to have resulted in significant observational features in the $Z$-$V_{Z}$  phase space \citep{Antoja2018, Laporte2019, Wang2019a}.    

According to the star-formation history modelling by  \cite{Ruiz2020}, the three passages of Sgr have triggered star formation, leading to a faster increase in [Fe/H] and jump in [$\alpha$/Fe] in the outer disk, which aligns with the observed flattened radial metallicity gradients in that region. Furthermore, Sgr's passage through the Galactic disk may also create a North-South metallicity asymmetry \citep{An2019}. However, that study's limited coverage of the Galactic disk and absence of age information prevented them from determining  the timing of Sgr's passage through the Galactic disk and the impact on its chemistry. 

In this paper, we seek to confirm whether or not the North-South metallicity asymmetry exists, and study its variations with spatial position and stellar ages using a red clump stellar sample selected from LAMOST DR8 and APOGEE DR16. 
This paper is organized as follows. Section 2 briefly
introduces the red clump stellar sample we employ. In Section 3, we present the main
results. A discussion of the possible origin associated with the first passage of Sgr through the Galactic disk is presented in Section 4, followed by a summary in Section 5.

\section{The red clump stellar samples}

In this paper, we employ the Galactocentric cylindrical coordinate system $(R, \Phi, Z)$, with $R$ representing the Galactocentric distance,
$\Phi$ increasing in the direction of Galactic rotation, $\Phi=0^{\circ} $ defined alone the anti-center Galactic direction,  and $Z$ representing the height from the disk mid-plane, positive towards the North Galactic pole. The Sun is assumed to be located at
($R_\sun, Z_\sun$) = (8.34, 0.025)\,kpc \citep{Juric2008, Reid2014}.

\subsection{Sample Selection}
Stars in the adopted red clump stellar sample are targeted by the LAMOST \citep{deng-legue, liu-lss-gac, yuan-lamost} and APOGEE \citep{GarcaPrez2016, Majewski2017, Jonsson2022} surveys.  Stars targeted by LAMOST are selected from a public catalog of $\sim 1$ million red giant branch and red clump stars with precise measurements of masses and ages ($\tau$) \citep{Wang2023}, derived from LAMOST DR8 low-resolution ($R \sim $1800) spectra with a neural network method using the LAMOST-Kepler giant stars as the training set.  For red clump stars, the typical uncertainties of stellar mass and age are 9\% and 24\%, respectively.  The metallicity of these stars comes from the  value-added catalog of LAMOST DR8, which provide estimates of stellar atmospheric parameters (effective temperature $T_{\mathrm{eff}}$, surface gravity $\log g$, metallicity ([Fe/H]$/$[M/H]), 
$\alpha$-element to metal abundance ratio [$\alpha$/M], carbon- and nitrogen-to-iron abundance ratios [C/Fe] and [N/Fe], and absolute magnitudes in 14 widely used photometric bands \citep{Wang2022}.  The typical uncertainties on metallicity are 0.05\,dex for stars with a spectral signal-to-noise ratio (SNR) larger than 50.  Stars targeted by APOGEE are selected from the APOGEE Red-Clump DR17 Catalog, which contains 50,387 red clump stars \citep{Bovy2014}. The stellar parameters are determined from the APOGEE Stellar Parameters and Chemical Abundances Pipeline \citep[ASPCAP;][]{Garca2016, Jonsson2020}; the precision of [M/H] is better than 0.1\,dex.  The ages of these APOGEE red clump stars is estimated using the machine learning technique XGBoost, trained on a high-quality dataset of 3060 red-giant and red-clump stars with asteroseismic ages observed by both APOGEE and Kepler \citep{Anders2023}. The typical uncertainty on stellar age is 17\%.  The distance measurements for these red clump stars are derived from \cite{Yu2025} and \cite{Bovy2014}. Due to the nature of red clump stars as standard candles, the typical distance uncertainties range from 5\% to 10\%, which is more accurate than the
measurements from Gaia parallaxes for stars beyond $\sim 5.0$\,kpc \citep{Huang2021}.

\begin{figure*}
\centering
\includegraphics[width=5.5in]{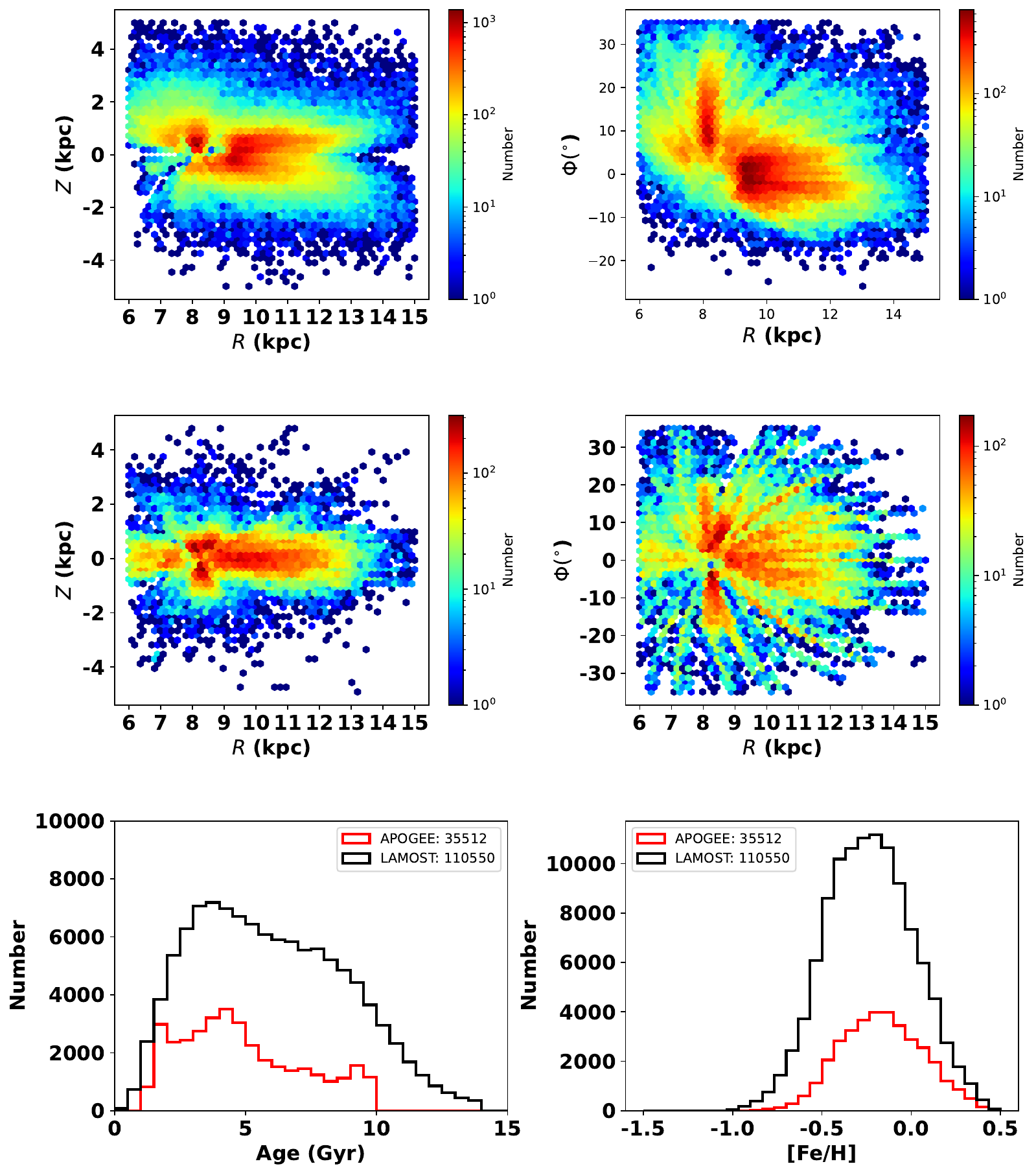}
\caption{The spatial distribution of the final adopted red clump stellar sample in the Galactic cylindrical coordinate with the first row displaying results of LAMOST: the $R$--$Z$ plane (left) and $R$--$\Phi$ plane (right). The second row presents the corresponding distributions for the APOGEE sample. In both cases, hexbin plots encode stellar number densities via the color bars. 
 These stars exhibit a distribution across the Milky Way spanning the ranges of $7$ to $14$\,kpc in Galactocentric radial distance $R$, $-5$ to $5$\,kpc in vertical distance from the Galactic plane $Z$, and $-35^\circ$ to $35^\circ$ in azimuthal angle $\Phi$. The bottom row shows age (left) and metallicity (right) distributions: black step histograms represent LAMOST, while red traces APOGEE.  The final red clump stellar sample encompasses a wide range of ages  from $0$ to $14$\,Gyr, and metallicities  ranging from $-1.5$ to $+0.5$.} 
\label{sample_distributions}
\end{figure*}

For red clump stars, we excluded stars with $\tau > 14$\,Gyr,  $|Z| > 5$\,kpc, $\rm [M/H]_{error} > 0.15$\,dex, and $\rm [\alpha/M]_{error} > 0.15$\,dex. For LAMOST red clump stars, we further discarded stars with spectral ${\rm SNR} < 30$, due to the age estimates of these stars having larger uncertainties.  In the end, a
total of 110,550 and 35,512 red clump stars are selected from LAMOST DR8 and  APOGEE DR17, respectively.  Figure\,\ref{sample_distributions} shows the spatial, metallicity, and age distributions of the final red clump sample. These stars exhibit a distribution across the Milky Way spanning the ranges of $7$ to $14$\,kpc in Galactocentric radial distance $R$, $-5$ to $5$\,kpc in vertical distance from the Galactic plane $Z$, and $-35^\circ$ to $35^\circ$ in azimuthal angle $\Phi$. Our red clump stellar sample encompasses a wide range of ages, from $0$ to $14$\,Gyr, and metallicities, ranging from $-1.5$ to $+0.5$.

\subsection{Systematic Offsets in the Age and Metallicity Distributions Between LAMOST and APOGEE Red Clump Stars}
\begin{figure}
\centering
\includegraphics[width=6.5in]{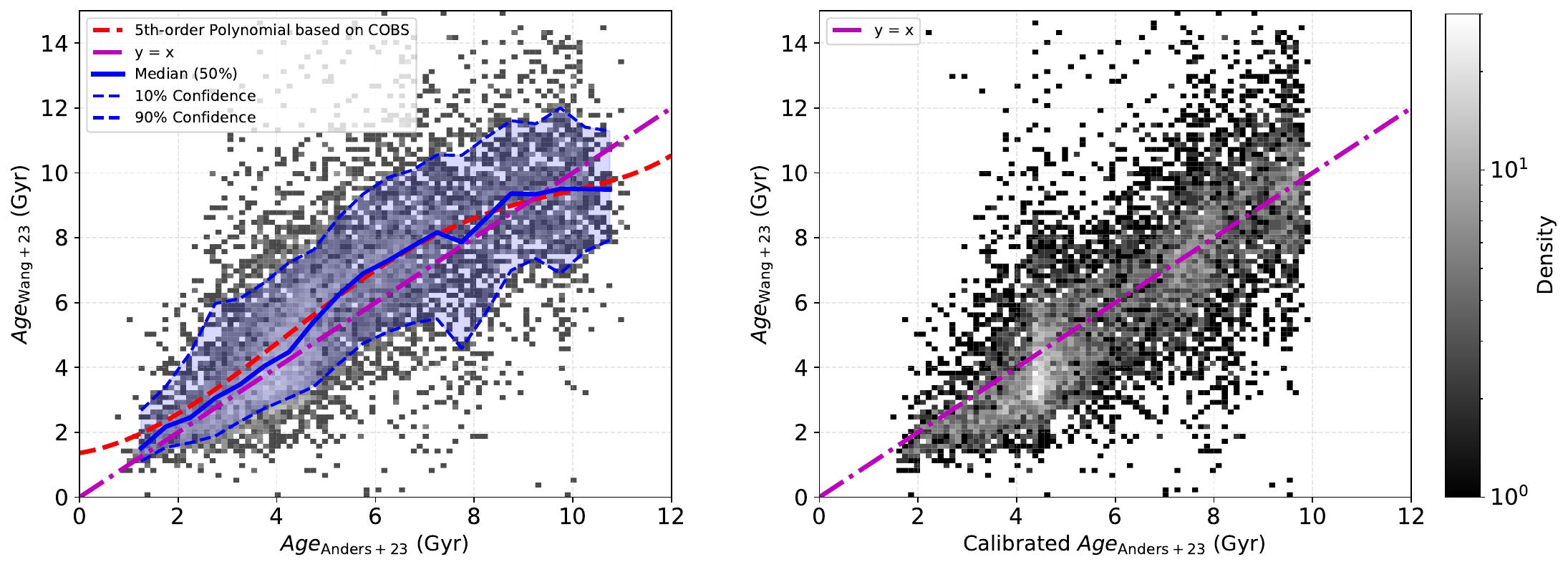}
\caption{Comparison of stellar age determinations between LAMOST and APOGEE red clump stars. Left panel:  Relationship between LAMOST and APOGEE ages derived from the constrained B-splines (COBS) nonparametric regression quantiles. The central 10th, 50th, and 90th percentile intervals are denoted by blue solid and dashed lines, while the polynomial fit is shown as red dashed lines. Right panel: Comparison of corrected ages. The violet dot-dashed line indicates the one to one relation. Point densities are encoded by grayscale values according to the color bar at right.}
\label{compare_age}
\end{figure}

As mentioned above, age estimates of red clump stars selected from LAMOST DR8 and APOGEE DR17 are determined using different methods by different groups, which may introduce systematic differences between the two surveys. Thus, we first compare the age estimates of red clump stars in common with both surveys to check this possibility. The comparison is shown in the left panel of Figure \ref{compare_age}.  We find that the ages of APOGEE red clump stars are higher at $\tau > 8$\,Gyr by 0.5-2\,Gyr, when compared to the LAMOST estimates.  
To correct for these systematic differences, we adopt a fifth-order polynomial derived from constrained B-splines (COBS) nonparametric regression quantiles \citep{Ng2015} to describe the trend in the age differences between LAMOST and APOGEE. This method integrates spline regression with likelihood-based knot selection and quantile regression to estimate the age relationship.
This small dependency is then mitigated by subtracting the fitted relation; the result is shown in the right panel of Figure \ref{compare_age}.
The calibrated age estimates for common red clump stars in LAMOST and APOGEE now exhibit close alignment with the one to one relation.
The metallicities of stars in LAMOST DR8 are estimated using a neural network method by taking the LAMOST-APOGEE common stars as the training set, thus there are no systematic differences between the metallicities of LAMOST and APOGEE, as shown in the Figure\,13 of \cite{Wang2022}.

\section{North-South Metallicity Asymmetries}
As shown in Figure\,\ref{sample_distributions}, the red clump stellar sample ranges from 7 to 14\,kpc in Galactocentric radial distance ($R$). The Galactic disk is known to exhibit significant radial metallicity gradients \citep{Mayor1976, Nordstrom2004, Huang2015, Xiang2015, Minchev2018,Wang2019}, which may affect our search for North-South metallicity asymmetries. Thus, to investigate the North-South metallicity asymmetry, we first subtracted the mean metallicity values in each $R$ bin; the radial metallicity gradient is effectively removed.

\subsection{North-South Metallicity Asymmetries of Stellar Populations at Different Azimuthal Angles}

We now investigate the North-South metallicity asymmetries of red clump stars at different azimuthal angles.  We divided stars from the red clump sample into six azimuthal angle bins: $-35^{\circ} < \Phi < -15^{\circ}$, $-15^{\circ} < \Phi < -5^{\circ}$, $-5^{\circ} < \Phi < 0^{\circ}$, $0^{\circ} < \Phi < 5^{\circ}$, $5^{\circ} < \Phi < 15^{\circ}$, and $15^{\circ} < \Phi < 35^{\circ}$. Within
each azimuthal angle bin, we further divided stars into different $R$--$Z$ bins, with a size of 0.6$\times$0.3\,(kpc)$^2$ for the $R \times Z$ dimensions. Bins containing fewer than three stars were discarded. Then we estimated the mean metallicity offset (the metallicity difference after subtracting the mean metallicity in each $R$ bin) in each spatial bin.  The radial metallicity gradients in $R$ are thus effectively removed, in order to better highlight the North-South metallicity asymmetries.  The original metallicity distributions and  in each $R$ bin and the subtracted radial mean metallicity  is shown in Figures \ref{mean_feh_z_phi} and \ref{mean_feh_z_age} of the Appendix.

Figure \ref{feh_rrzz_final_phi} shows the metallicity-offset distributions in the  $R$--$Z$ plane at different azimuthal angle $\Phi$.  In the inner disk region ($R < 11$\,kpc), the metallicity-offset distribution exhibits a clear mirror symmetry with respect to the Galactic plane ($Z=0$\,kpc). In contrast, in the outer disk region ($R > 11$\,kpc), an examination of the metallicity-offset distribution reveals a distinct North-South asymmetry in the azimuthal angle range of $-5^\circ$ to $35^\circ$ (Fig. 3c, d, e, f). Specifically, the metallicities in the Northern Hemisphere exhibit higher values compared to the Southern 
Hemisphere, with an approximate difference in the range $0.0$--$0.15$\,dex. However, the metallicity-offset distribution within the azimuthal angle range of $-35^\circ$ to $-5^\circ$ (Fig. 3a, b) shows no significant North-South asymmetry, even at $R > 11$\,kpc.  

\begin{figure*}
\centering
\includegraphics[width=6.5in]{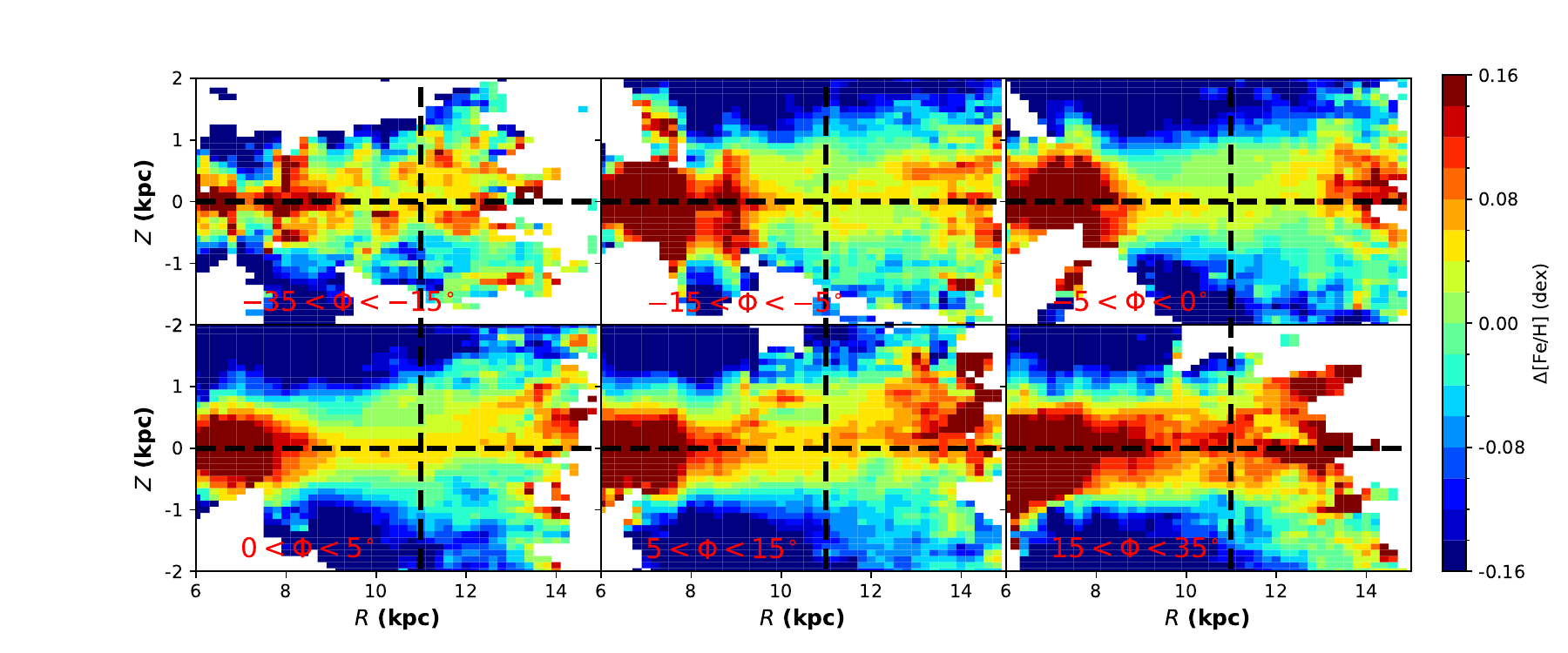}
\caption{Metallicity-offset distributions of our red clump stellar sample in  the $R$-$Z$ plane across different azimuthal angles.  The colors represent the metallicity offsets (the mean metallicity with radial metallicity gradients removed), coded in the color bar to the right. White bins are those containing fewer than three stars.}
\label{feh_rrzz_final_phi}
\end{figure*}

To quantitatively assess the North-South asymmetry in the outer disk ($R > 11$\,kpc), we calculate the mean metallicity within different $Z$ bins, and analyze the variations in metallicity with respect to $Z$, as depicted in Fig.\,4a. Figures 4b and 4c illustrate the metallicity asymmetry observed in the outer disc. Here, the metallicity asymmetry, denoted as $A$, 

\begin{equation}
A = \frac{[\mathrm{Fe/H}]_\mathrm{North} - [\mathrm{Fe/H}]_\mathrm{South}}{[\mathrm{Fe/H}]_\mathrm{North} + [\mathrm{Fe/H}]_\mathrm{South}} \\
\end{equation}
, where $\rm [Fe/H]_{North}$ and $\rm [Fe/H]_{South}$ represent the median metallicity of stars in the Northern and Southern hemispheres, respectively. We find that, for populations within the range of $-5^\circ$ to $15^\circ$, the median metallicity at $Z > 0$\,kpc is slightly higher than that at $Z < 0$\,kpc, resulting in a positive metallicity asymmetry value. However, for the azimuthal angle range of $-15^\circ < \Phi < -5^\circ$, the metallicity asymmetry is approximately zero. When considering $15^\circ < |\Phi| < 35^\circ$, the metallicity asymmetries exhibit larger errors and oscillate around zero. We conclude that the metallicity asymmetry is a noteworthy localized characteristic in the Galactic disk.

\begin{figure}
\centering
\includegraphics[width=6.5in]{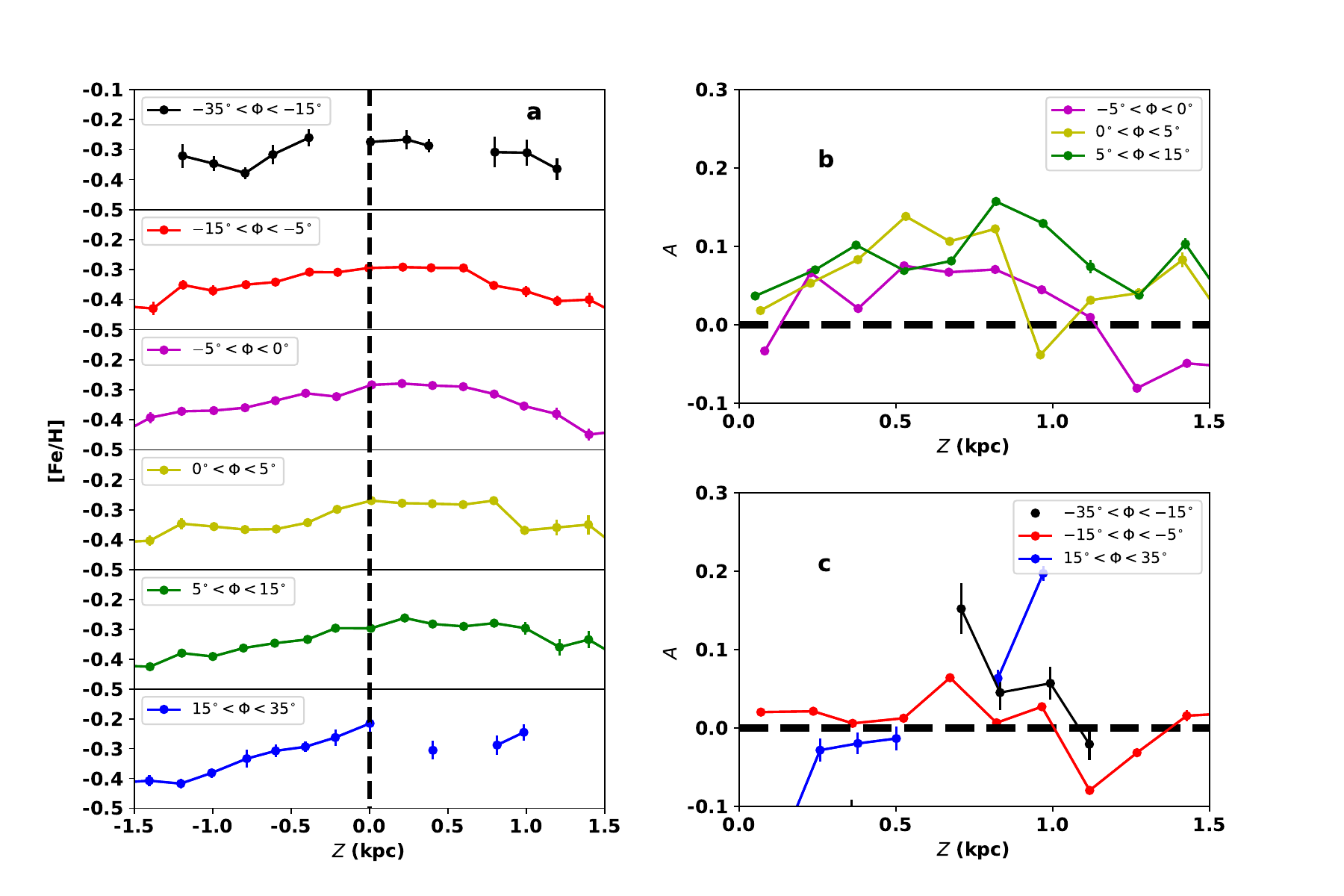}
\caption{North-South metallicity asymmetries in the outer disk ($R > 11\,$kpc). 
Median metallicity variations with respect to $Z$ for stellar populations at different azimuthal angles. 
Panel b: Variations of the metallicity asymmetry ($A$, see text), as a function of $|Z|$, for stellar populations over $-5^{\circ} < \Phi < 15^{\circ}$. The metallicity asymmetries can be found in this panel. 
Panel c: Variations of metallicity asymmetry $A$, as a function of $|Z|$, for stellar populations within $-35^{\circ} < \Phi < -5^{\circ}$ and $15^{\circ} < \Phi < 35^{\circ}$.  The metallicity asymmetries shown in this panel are not significant.  }
\label{feh_z_phi} 
\end{figure}

\vskip 1cm
\subsection{North-South Metallicity Asymmetries for Different Age Stellar Populations}

\begin{figure*}
\centering
\includegraphics[width=6.5in]{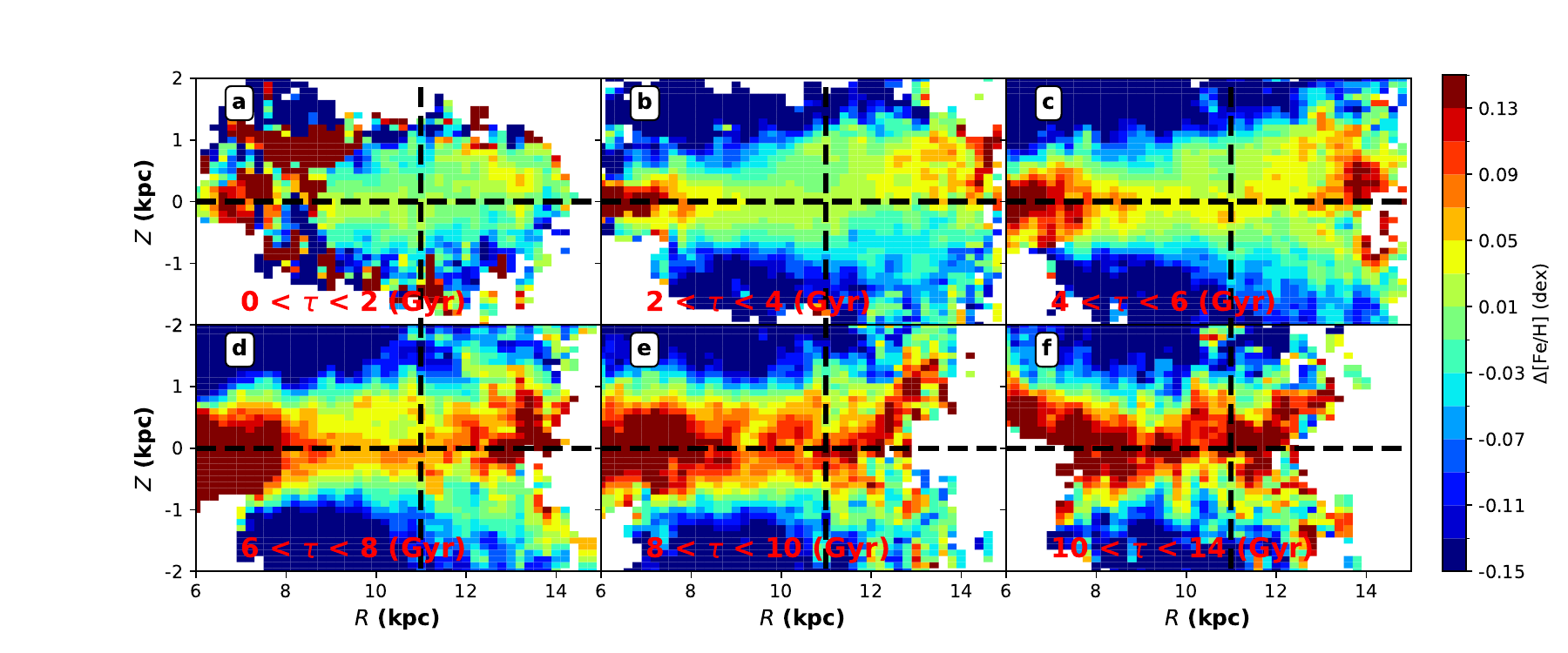}
\caption{Metallicity-offset distributions for various stellar populations of different ages within the azimuthal range of $-5^{\circ}$ to $15^{\circ}$, as coded in the color bar to the right. }
\label{feh_rrzz_final}
\end{figure*}

The metallicity exhibits its greatest North-South asymmetry within the azimuthal range of $-5^\circ < \Phi < 15^\circ$, as demonstrated in Figures \ref{feh_rrzz_final_phi} and \ref{feh_z_phi}. In the case of these stars, we focused on investigating the metallicity asymmetry within the most heterogeneous stellar populations by charting the distributions of metallicity among various stellar populations ($0 < \tau < 2$\,Gyr, $2 < \tau < 4$\,Gyr, $4 < \tau < 6$\,Gyr, $6 < \tau < 8$\,Gyr, $8 < \tau < 10$\,Gyr and $10 < \tau < 14$\,Gyr), illustrated in Figure \ref{feh_rrzz_final}.

\begin{figure}
\centering
\includegraphics[width=6.5in]{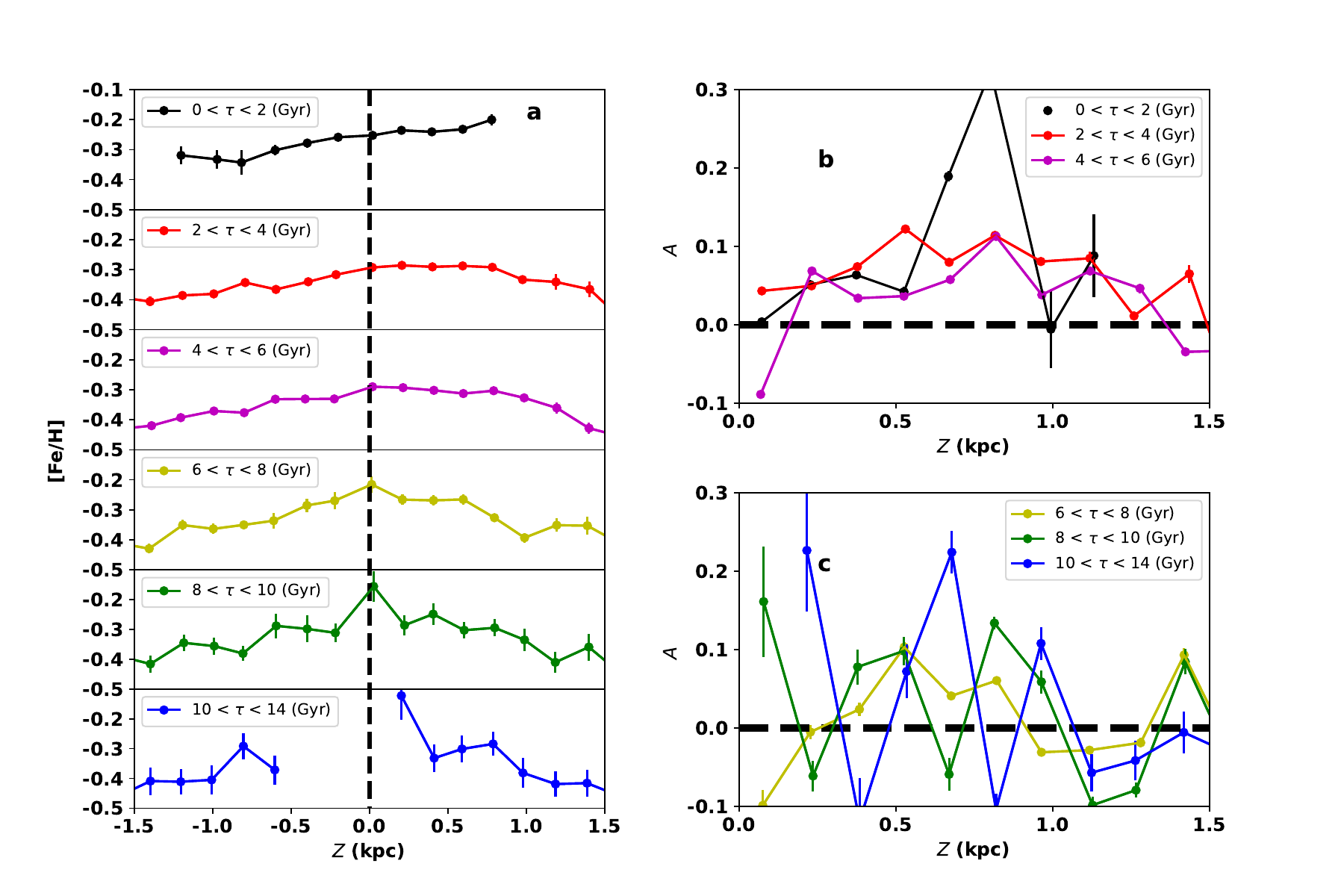}
\caption{The North-South metallicity asymmetry for $R > 11$\,kpc and $-5^{\circ} < \Phi < 15^{\circ}$ across distinct stellar age groups. 
Panel a: Metallicity variations, with respect to $Z$, for different stellar populations. 
Panel b: Variations of the metallicity asymmetry $A$, as a function of $|Z|$, for younger stellar populations. The metallicity asymmetries are found in this panel. 
Panel c: Variations of the metallicity asymmetry $A$, as a function of $|Z|$ for older stellar populations. The metallicity asymmetries shown in this panel are not significant.  }

\label{feh_z_age}
\end{figure}

In the outer disk ($R > 11$\,kpc),  we can see a North-South metallicity asymmetry for all these mono-age stellar populations, especially for the younger stellar populations with age $\tau <6$\,Gyr.  Hemisphere exceeds that in the Southern Hemisphere by a margin of 0.0--0.15\,dex for younger stellar populations.  The North-South metallicity asymmetry of the older stellar populations ($\tau > 8$\,Gyr) are not so significant than that of younger ones.  This reduced significance likely stems from larger age uncertainties and fewer stars in the outer disk.  Additionally, older stellar populations are kinematically hotter than younger ones \citep{Das2024}, which may have partially erased the metallicity asymmetry \citep{Chequers2018}. 
 Figure \ref{feh_z_age} shows the median metallicity and metallicity asymmetry $A$, as functions of $Z$, for different stellar populations at $R > 11$\,kpc. 
From inspection of Figure 5a, it is evident that younger stellar populations exhibit a notable North-South metallicity asymmetry, with stars in the Northern Hemisphere exhibiting a higher metallicity offset compared to their Southern Hemisphere counterparts. This phenomenon is less pronounced in older stellar populations.  Specifically, Figure 5b illustrates that, for the younger stellar population, the metallicity asymmetry is prominently positive, while Figure 6c indicates that this metallicity asymmetry tends to converge to around zero at large amplitudes for the older stellar populations. 

\vskip 2cm
\section{Origin of the observed North-South metallicity asymmetry}
The metallicity-offset distributions in the $R$-$Z$ plane as traced by the LAMOST and APOGEE red clump stellar sample presented above suggest the presence of North-South metallicity asymmetries, which depend on the Galactic radius, azimuthal angles, and stellar ages.  
All stellar populations (especially for young stellar population of $\tau < 6$\,Gyr) exhibit North-South metallicity asymmetries, which are evident in the outer disk ($R > 11$\,kpc) in the direction toward the anti-Galactic center ($-5^{\circ} < \Phi < 15^{\circ}$); the metallicity of stars in the Northern Galactic Hemisphere is higher than that of stars in the Southern Galactic Hemisphere. Similarly, \cite{An2019} found a metallicity asymmetry in the Galactic disk based on the metallicity distributions of stars from SDSS DR14, using photometric metallicity estimates.  They found that the metallicity asymmetry exhibits wave-like oscillations, consistent with the density asymmetry reported by \citet{Widrow2012}, \citet{Yanny2013}, and \citet{Bennett2019}. In contrast, we did not find a clear wave-like oscillation in the metallicity asymmetry, as we focused on the outer disk rather than on the Solar Neighborhood. 

One possible origin of the North-South metallicity asymmetry is a perturbation caused by the passage of a dwarf galaxy through the Galactic disk.  When the dwarf galaxy passes through the Galactic plane, it may produce the observed density and kinematic North-South asymmetry.  This passage could also induce a transient increase in the star formation rate (SFR) within a relatively short time scale of approximately 1--2\,Gyr, with the enhanced SFR mainly seen in the outer disk of the host galaxy \citep{Annem2023}.   If a dwarf galaxy passes through the disk from South to North, the gas and stars in the Galactic disk might become compressed in this direction. The Galactic disk plane may have experienced a localized upward displacement, which contribute to the observed North-South metallicity asymmetry, particularly in the outer disk regions.  Besides,  the compressed gas from South to North will produce higher SFR in the North and finally giving rise to the observed North-South metallicity asymmetry.  A minor (yet quite massive) merger is expected to have deposited $\sim$ 50\% of its gas to its host,  thus diluting the disk in [Fe/H] \citep{Tepper2018}.  As the dwarf galaxy moves through the disk from south to north, a larger quantity of gas accreted from the dwarf will accumulate in the South. As a result, the stars that form in this region will be more metal-poor than those formed in the North.   

We can provide some constraints on the passage event in this scenario.  This passage leads to enhanced star formation rates and significant displacement in gas position, primarily observed in the outer disk and the corresponding azimuthal angle of the dwarf galaxy's passage.  According to the results in section 3,  this asymmetry is most obvious in the outer disk ($R > 11$\,kpc), which can constrain the Galactic radius of the dwarf galaxy's passage. In the outer disk, the North-South metallicity asymmetry is most obvious in the direction of the anti-Galactic center ($-5^{\circ} < \Phi < 15^{\circ}$) .
We conclude that such a dwarf galaxy traversed the Galactic disk within the azimuthal range of $-5^{\circ} < \Phi < 15^{\circ}$, consistent with the result  of  \cite{Wang2019a}.

An alternative explanation for the observed North-South metallicity asymmetry is that the Galactic disk is being affected by tidal forces from a more recent interaction with a massive halo substructure. In this scenario, stars in the Galactic disk becomes phase space warped in the $Z$--$V_{z}$ plane following the passage of a dwarf galaxy, which will induce a local metallicity perturbation from the mean vertical metallicity gradient \citep{An2019}.  The most likely candidate for such a dwarf galaxy is also the Sagittarius dwarf galaxy. 
According to the simulation of \cite{Chequers2018}, the bending waves induced by subhalos are more vigorously excited in the thin disk than the thick disk. They also suggest that the perturbation effects are strongest in the outer regions of the disk.   Our metallicity asymmetries are strongest for younger stellar populations and in the outer disk, which is consistent with the prediction of \cite{Chequers2018}. 
Of course, the less pronounced metallicity asymmetry of older stellar populations may also  be attributed to the lower accuracy in age determinations and the  fewer   number of stars in our sample for older stellar populations. In the future, greater precision in age measurements and broader spatial coverage of older stars in the Galactic disk could help us address this question. 

The observed North-South metallicity asymmetry could also reflect the chemical characteristics of the Galactic warp. Previous studies on the Galactic warp indicate that its line of node is approximately at $\Phi \sim 12.5^{\circ}$, in which direction the warp is most neglible \citep{Lixin2020}. However, this finding contradicts our results, which show that the most pronounced metallicity asymmetry falls within the range of $-5^{\circ} < \Phi < 15^{\circ}$. 

In conclusion, the North-South metallicity asymmetry observed may be a result of the recent passage and tidal forces of the Sagittarius dwarf galaxy.  The less pronounced metallicity asymmetry in older stellar populations requires larger sample sizes with broader spatial coverage and higher age-determination accuracy to confirm its physical reality.

\section{Summary}
In this work, we use the LAMOST and APOGEE red clump stellar sample with accurate age and metallicity estimates to investigate the North-South metallicity asymmetry of different stellar populations across the Galactic disk over the ranges of $6 < R < 15$\,kpc, $|Z| < 2.0$\,kpc and $-35^{\circ} < \Phi < 35^{\circ}$. 

We first considered the metallicity-offset distributions in the $R$-$Z$ plane, and found that the metallicity of stars in the Northern Galactic Hemisphere is higher than that of stars in the Southern Galactic Hemisphere 
in the outer disk ($R > 11$\,kpc), especially toward the anti-Galactic center ($-5^{\circ} < \Phi < 15^{\circ}$) direction, giving rise to a  North-South metallicity asymmetry.  In the inner disk ($R < 11$\,kpc), the metallicity-offset distributions in the Northern and Southern Galactic Hemispheres are, in contrast, almost identical.

In the outer disk and the anti-Galactic center direction, we then mapped out the metallicity distributions in the $R$-$Z$ plane for different stellar populations. We found that the younger stellar populations with ages $\tau < 6$\,Gyr manifest a North-South metallicity asymmetry. Specifically, the metallicity of stars in the Northern Hemisphere is higher than that in the Southern Hemisphere by a margin of 0.0--0.15\,dex.  The metallicity asymmetry of older stellar populations ($\tau > 8$\,Gyr) is less pronounced than that of younger stellar populations, which may stems from three factors: larger age uncertainties, fewer stars in the outer disk, and the kinematically hotter nature of older populations. 

The perturbation created by the most recent passage through the Galactic disk and tidal force of the Sagittarius dwarf galaxy  may lead to the observed North-South metallicity asymmetry. To replicate this observed metallicity asymmetry accurately, detailed simulations will be required in the future. 



\begin{acknowledgments}
The authors would like to thank the referee and editor for their insightful comments that improved the quality of this manuscript. This work was funded by the National Key R\&D Program of China
(No. 2019YFA0405500) and the National Natural Science Foundation of China (NSFC Grant Nos. 11988101, 12422303, 11833006, 11973001, 12173007, 12422303, and 12273053).
Guoshoujing Telescope (the Large Sky Area Multi-Object Fiber Spectroscopic
Telescope LAMOST) is a National Major Scientific Project built by the Chinese Academy of Sciences.
Funding for the project has been provided by the National Development and Reform Commission.
LAMOST is operated and managed by the National Astronomical Observatories, Chinese Academy of
Sciences. The LAMOST FELLOWSHIP is supported by Special Funding for Advanced Users, budgeted and administrated by Center for Astronomical Mega-Science, Chinese Academy of Sciences (CAMS). Supported by High-performance Computing Platform of Peking University.  T.C.B. acknowledges partial support from Physics Frontier Center/JINA Center for the Evolution of the Elements (JINA-CEE), and OISE-1927130: The International Research Network for Nuclear Astrophysics (IReNA), awarded by the US National Science Foundation (NSF).
\end{acknowledgments}

\bibliography{metal-asym.bib}{}
\bibliographystyle{aasjournal}

\begin{figure}
\centering
\includegraphics[width=6.5in]{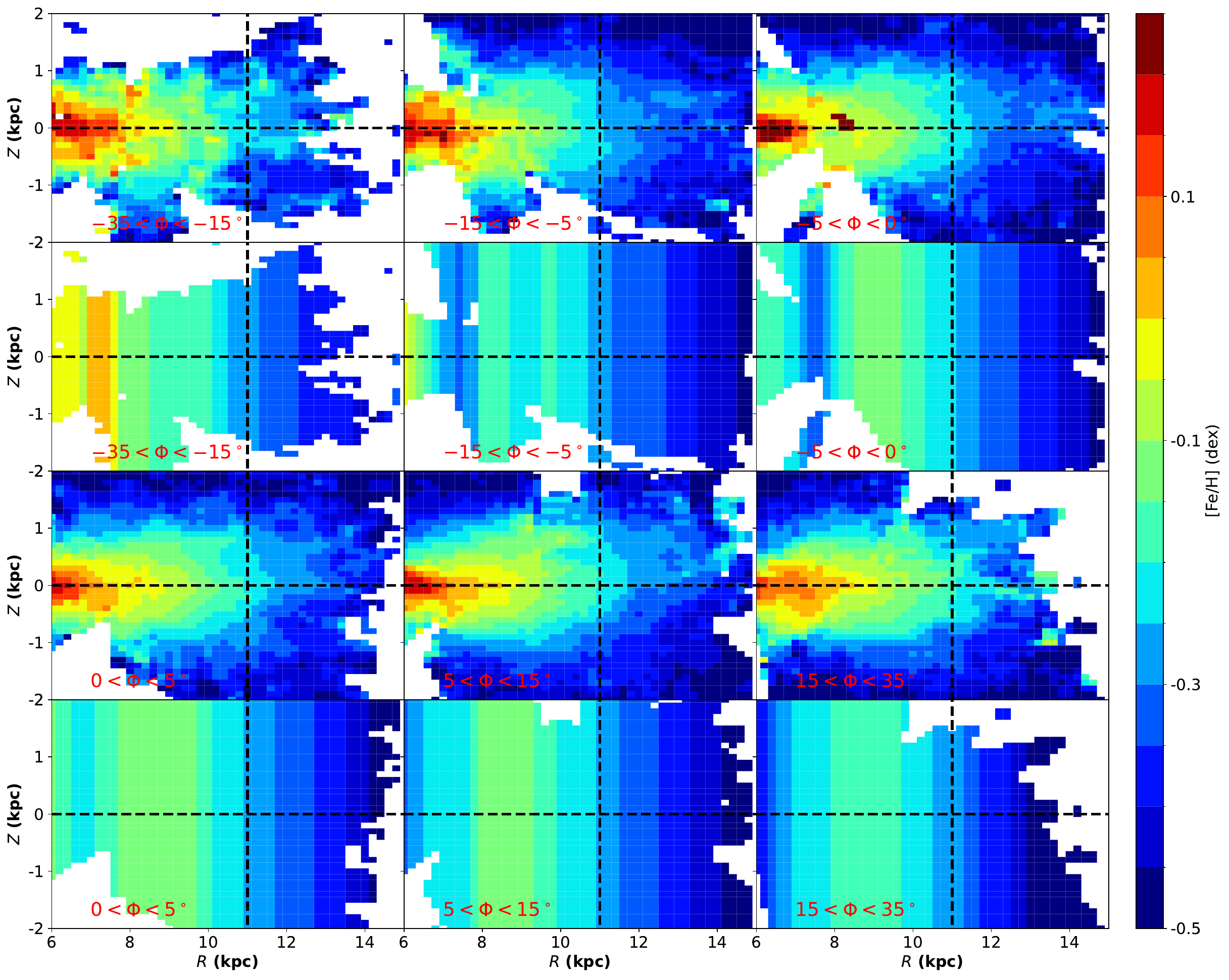}
\caption{The distributions of original metallicity (the first and third rows) and the subtracted radial mean metallicity  (the second and fourth rows) of our red clump stellar sample in  the $R$-$Z$ plane across different azimuthal angles.  The colors represent the metallicity, coded in the color bar to the right. White bins are those containing fewer than three stars.}
\label{mean_feh_z_phi}
\end{figure}
\begin{figure}
\centering
\includegraphics[width=6.5in]{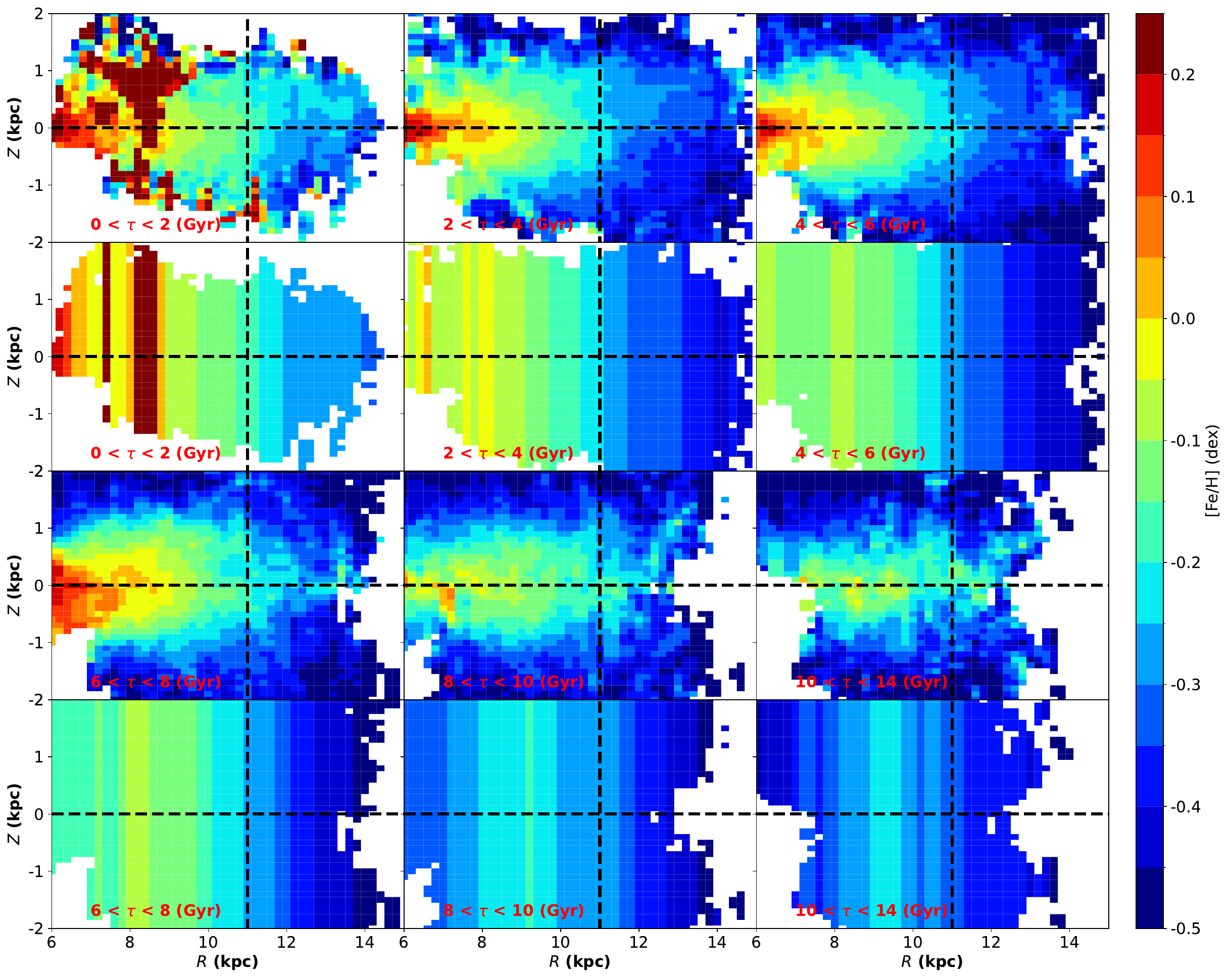}
\caption{Similar to Figure \ref{mean_feh_z_phi}, but showing results for separate age stellar populations.}
\label{mean_feh_z_age}
\end{figure}

\appendix
Here we present: (i) the original metallicity distributions, and (ii) the radial mean metallicity within each $R$ bin across distinct azimuthal angles (Figure\,\ref{mean_feh_z_phi})and for separate stellar age populations (Figure\,\ref{mean_feh_z_age}).

\end{document}